\documentclass[12pt,preprint,fleqn]{aastex}
\usepackage{graphicx}
\slugcomment{Accepted for publication in {\it Astroparticle Physics}}

\begin{document}

\title{Constraints on the Geometry of the VHE Emission in
LS 5039 from Photon-Photon Deabsorption}

\author{Markus B\"ottcher\altaffilmark{1}
}

\altaffiltext{1}{Astrophysical Institute, Department of Physics 
and Astronomy, Ohio University, Athens, OH 45701, USA}

\begin{abstract}
Evidence for orbital modulation of the very high energy (VHE) 
$\gamma$-ray emission from the high-mass X-ray binary and 
microquasar LS~5039 has recently been reported by the HESS
collaboration. The observed flux modulation was found to go
in tandem with a change in the GeV -- TeV spectral shape,
which may partially be a result of $\gamma\gamma$ absorption
in the intense radiation field of the massive companion star. 
However, it was suggested that $\gamma\gamma$ absorption effects
alone can not be the only cause of the observed spectral variability
since the flux at $\sim 200$~GeV, which is near the minimum of the 
expected $\gamma\gamma$ absorption trough, remained essentially 
unchanged between superior and inferior conjunction of the binary
system. In this paper, a detailed parameter study of the $\gamma\gamma$
absorption effects in this system is presented. For a range of plausible
locations of the VHE $\gamma$-ray emission region and the allowable 
range of viewing angles, the de-absorbed, intrinsic VHE $\gamma$-ray
spectra and total VHE photon fluxes and luminosities are calculated
and compared to luminosity constraints based on Bondi-Hoyle limited
wind accretion onto the compact object in LS~5039. Based on these 
arguments, it is found that (1) it is impossible to choose the viewing 
angle and location of the VHE emission region in a way that the intrinsic 
(deabsorbed) fluxes and spectra in superior and inferior conjunction 
are identical; consequently, the intrinsic VHE luminosities and 
spectral shapes must be fundamentally different in different 
orbital phases, (2) if the VHE luminosity is limited by wind 
accretion from the companion star and the system is viewed at 
an inclination angle of $i \gtrsim 40^o$, the emission is most
likely beamed by a larger Doppler factor than inferred from the 
dynamics of the large-scale radio outflows, (3) the still poorly 
constrained viewing angle between the line of sight and the jet 
axis is most likely substantially smaller than the maximum of 
$\sim 64^o$ inferred from the lack of eclipses.
(4) Consequently,
the compact object is more likely to be a black hole rather than
a neutron star. (5) There is a limited range of allowed configurations
for which the expected VHE neutrino flux would actually
anti-correlate with the observed VHE $\gamma$-ray emission.
If hadronic models for the $\gamma$-ray production in LS~5039 apply,
a solid detection of the expected VHE neutrino flux and its orbital
modulation with km$^3$ scale water-Cherenkov neutrino detectors might 
require the accumulation of data over more than 3 years.
\end{abstract}


\section{\label{intro}Introduction}

\begin{figure}[t]
\includegraphics[width=14cm]{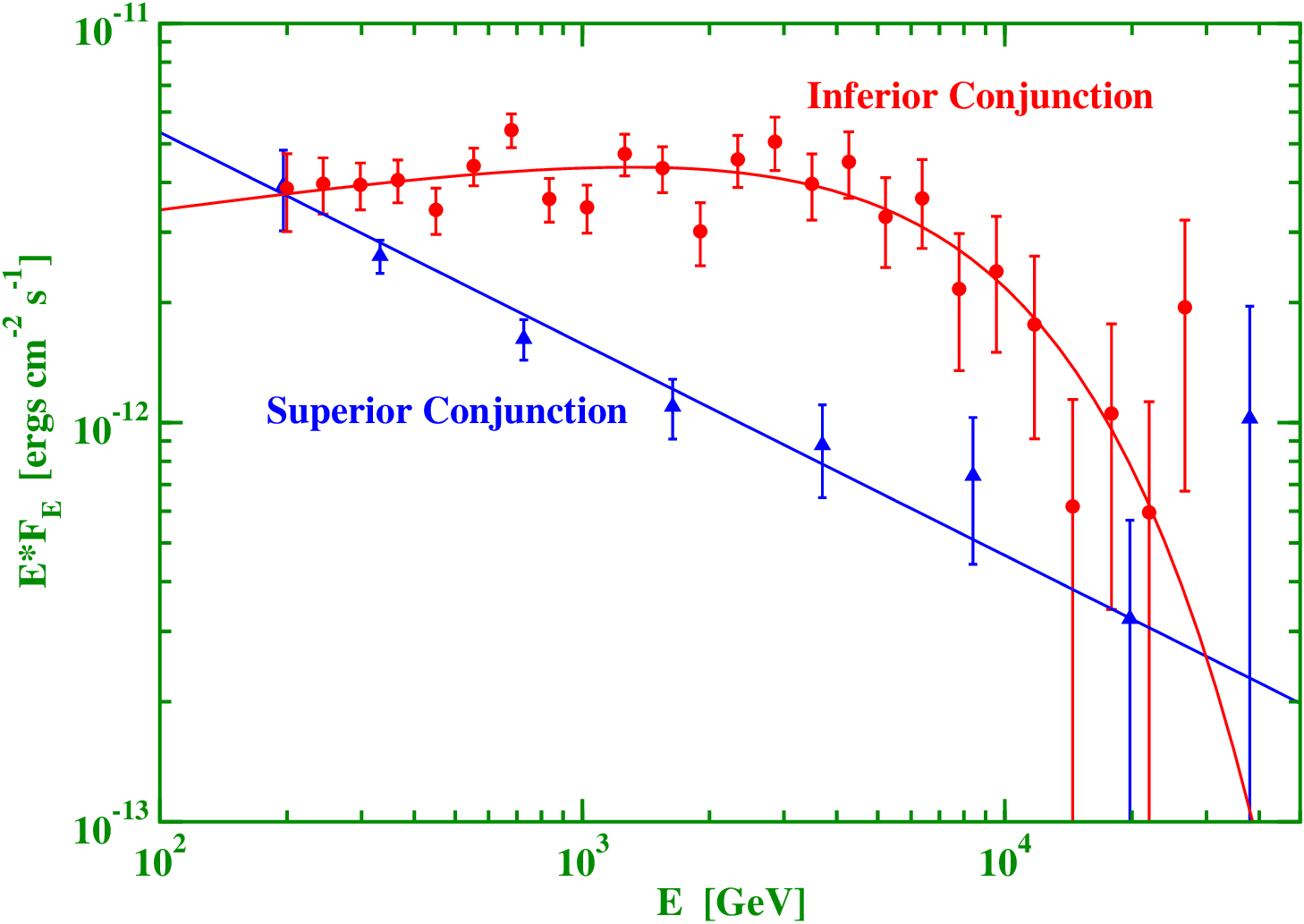}
\caption{VHE $\gamma$-ray spectra of LS~5039 around superior (blue 
triangles) and inferior (red circles) conjunction 
\citep[from][]{aharonian06b}. The curves indicate the best-fit straight
power-law (superior) and exponentially cut-off power-law (inferior)
representations of the spectra, on which the analysis in this paper
is based.}
\label{observed_spectrum}
\end{figure}

The recent detections of VHE ($E \gtrsim 250$~GeV) $\gamma$-rays 
from the high-mass X-ray binary jet sources LS 5039 with the High 
Energy Stereoscopic System \citep[HESS;][]{aharonian05}
and LS~I~+63$^o$303 with the Major Atmospheric Gamma-Ray Imaging 
Cherenkov Telescope \citep[MAGIC][]{albert06} establish these
sources (termed ``microquasars'' if they are accretion powered)
as a new class of $\gamma$-ray emitting sources. These results 
confirm the earlier tentative identification
of LS~5039 with the EGRET source 3EG~J1824-1514 
\citep{paredes00} and LSI~$61^o303$ with the COS~B source 2CG~135+01
\citep{gregory78,taylor92} and the EGRET source 3EG~J0241+6103 
\citep{kniffen97}.
Both of these objects show evidence 
for variability of the VHE emission, suggesting an association with 
the orbital period of the binary system. In the case of LS~I~+63$^o$303, 
the association with the orbital period is not yet firmly established 
since the MAGIC observations covered only a few orbital periods, and 
the orbital period \citep[$P = 26.5$~d;][]{gregory02} is very close 
to the siderial period of the moon, which also sets a natural windowing
period for VHE observations \citep{albert06}. In contrast, the HESS
observations of LS~5039 provide rather unambiguous evidence for an 
orbital modulation of both the VHE $\gamma$-ray flux and spectral shape 
with the orbital period of $P = 3.9$~d \citep{aharonian06b}. Specifically,
\cite{aharonian06b} found that between inferior conjunction (i.e., the
compact object being located in front of the companion star), the VHE
$\gamma$-ray spectrum could be well fitted with an exponentially cut-off,
hard power-law of the form $\Phi_E \propto E^{-1.85} \, e^{-E/E_0}$
whith a cut-off energy of $E_0 = 8.7$~TeV. In contrast, the VHE spectrum
at superior conjunction is well represented by a pure power-law 
($\Phi_E \propto E^{-2.53}$) with a much steeper slope, but identical 
differential flux at $E_{\rm norm} \approx 200$~GeV. The original data 
from \cite{aharonian06b}, together with these spectral fit functions, 
are shown in Fig. \ref{observed_spectrum}.

A variety of different models for the high-energy emission from 
high-mass X-ray binaries have been suggested. These range from 
high-energy processes in neutron star magnetospheres 
\citep[e.g.,][]{moskalenko93,moskalenko94,bednarek97,bednarek00,chernyakova06,dubus06b},
via models with the inner regions of microquasar jets being the primary high-energy
emission sites \citep[e.g.][]{romero03,bp04,bosch05a,gupta06,db06,gb06}
and interactions of microquasar jets with the ISM \citep{bosch05b},
to models involving particle acceleration in shocks produced by colliding 
stellar winds \citep[e.g.,][]{reimer06}. In addition to involving a variety
of different leptonic and hadronic emission processes, these models also
imply vastly different locations of the $\gamma$-ray production site with
respect to the compact object in LS~5039 and the companion star. However,
independent of the emission mechanism responsible for the VHE $\gamma$-rays,
the intense radiation field of the high-mass (stellar type O6.5V) companion 
will lead to $\gamma\gamma$ absorption of VHE $\gamma$-rays in the $\sim 100$~GeV
-- TeV photon energy range. The characteristic features of the $\gamma\gamma$
absorption of VHE $\gamma$-rays by the companion star light in LS~5039
have been investigated in detail by \cite{bd05} and \cite{dubus06a}. It
was found that, if the $\gamma$-ray emission originates within a distance
of the order of the orbital separation of the binary system ($s \sim 2 \times
10^{12}$~cm), this effect should lead to a pronounced $\gamma\gamma$ absorption
trough, in particular near superior conjunction, while $\gamma\gamma$ absorption
tends to be almost negligible near inferior conjunction. In LS~5039, the minimum
of the absorption trough is expected to be located around $E_{\rm min} \sim
300$~GeV and should shift from higher to lower energies as the orientation of 
the binary system changes from inferior to superior conjunction. 

Most of the relevant parameters of LS~5039, except for the inclination angle
$i$ of the line of sight with respect to the normal to the orbital plane, are 
rather well determined (see \S \ref{parameters}). This allows for a detailed 
parameter study of the $\gamma\gamma$ absorption effects, leaving the 
inclination angle and the distance of the VHE emission site from the compact 
object as free parameters. As will be shown in \S \ref{cascades}, the effect
of electromagnetic cascades will lead to a re-deposition of the absorbed
$\gtrsim 100$~GeV luminosity almost entirely at photon energies $E \lesssim
100$~GeV. For that reason, one can easily correct for the $\gamma\gamma$
absorption effect at energies $E \gtrsim 100$~GeV by multiplying the observed
fluxes by a factor $e^{\tau_{\gamma\gamma}(E)}$ (where $\tau_{\gamma\gamma}(E)$
is the $\gamma\gamma$ absorption depth along the line of sight) in order 
to find the intrinsic, deabsorbed VHE spectra for any given choice of $i$
and the height $z_0$ of the VHE $\gamma$-ray emission site above the compact
object. Results of this procedure will be presented in \S \ref{deabsorbed}.
In \S \ref{constraints}, these results will then be used to estimate the 
total flux and luminosity in VHE $\gamma$-rays, which can be compared to
limits on the available power under the assumption that the VHE emission
is powered by Bondi-Hoyle limited wind accretion onto the compact object. 
This leads to important constraints on the location of the VHE emision
site and the geometry of the system, which will be discussed in \S 
\ref{summary}.

\section{\label{parameters}Parameters of the LS~5039 System}

LS~5039 is a high-mass X-ray binary in which a compact object is
in orbit around an O6.5V type stellar companion with a mass of
$M_{\ast} = 23 \, M_{\odot}$, a bolometric luminosity of $L_{\ast} 
= 10^{5.3} \, L_{\odot} \approx 7 \times 10^{38}$~ergs~s$^{-1}$
and an effective surface temperature of $T_{\rm eff} = 39,000 \, ^o$K.
The mass function of the system is $f(M) = (M_{\rm c.o.} \sin i)^3
/ (M_{\rm c.o.} + M_{\ast})^2 \approx 5 \times 10^{-3} \, M_{\odot}$. 
The binary orbit has an eccentricity of $e = 0.35$ and an orbital 
period of $P = 3.9$~d. The inclination angle $i$ is only poorly 
constrained in the range $13^o \lesssim i \lesssim 64^o$. Under 
the assumption of co-rotation of the star with the orbital motion, 
a preferred inclination angle of $i = 25^o$ could be inferred, 
leading to a compact-object mass of $M_{\rm c.o.} = 3.7^{+1.3}_{-1.0} 
\, M_{\odot}$ \citep{casares05}. However, since there is no clear 
evidence for co-rotation, the inclination angle is left as a free 
parameter in this analysis. Within the entire range of allowed 
values, $13^o \le i \le 64^o$, the mass of the compact object 
is substantially smaller than the mass of the companion, so 
that the orbit can very well be approximated by a stationary 
companion star, orbited by the compact object for our purposes. 
The orbit has a semimajor axis of length $a = 2.3 \times 10^{12}$~cm, 
and the projection of the line of sight onto the orbital plane 
forms an angle of $\approx 45^o$ with the semimajor axis 
\citep[see, e.g., Fig. 4 of][]{aharonian06b}. Consequently,
the orbital separation at superior conjunction is found to be
$s_{\rm s.c.} = 1.6 \times 10^{12}$~cm, while at inferior conjunction,
it is $s_{\rm i.c.} = 2.7 \times 10^{12}$~cm. 

The mass outflow rate in the stellar wind of the companion has been 
determined as $\dot M_{\rm wind} \approx 10^{-6.3} M_{\odot}$/yr, and 
the terminal wind speed is $v_{\infty} \approx 2500$~km~s$^{-1}$ \citep{mg02}.
EVN and MERLIN observations of the radio jets of LS~5039 suggest a 
mildly relativistic flow speed of $\beta \sim 0.2$ on the length 
scale of several hundred AU \citep{paredes02}. This would correspond
to a bulk Lorentz factor of the flow of $\Gamma \approx 1.02$.
However, it is plausible to assume that near the base of the jet,
where the VHE $\gamma$-ray emission may arise (according to some
of the currently most actively discussed modeles), the flow may 
have a substantially higher speed. Therefore, we will also consider
bulk flow speeds of $\Gamma \sim 2$, more typical of the jet speeds
of other Galactic microquasar jets. 

For the analysis in this paper, we use a source distance of
$d = 2.5$~kpc, corresponding to $4 \pi d^2 = 7.1 \times 10^{44}$~cm$^2$
\citep{casares05}.

Inferred compact object masses and Doppler boosting factors $D = 
(\Gamma [1 - \beta \cos i])^{-1}$ for representative values of 
$i = 20^o$, $40^o$, and $60^o$ and $\Gamma = 1.02$ and $\Gamma = 2$
are listed in Table \ref{parameter_table}.

\section{\label{cascades}The Role of Electromagnetic Cascades}

The absorption of VHE $\gamma$-ray photons photons will lead to the 
production of relativistic electron-positron pairs, which will lose 
energy via synchrotron radiation and inverse-Compton scattering on 
starlight photons, and initiate synchrotron or inverse-Compton supported
electromagnetic cascades. For photon energies of $E_{\gamma} \gtrsim 
100$~GeV, the produced pairs will have Lorentz factors of $\gamma
\gtrsim 10^5$. Given the surface temperature of the companion star
of $T_{\rm eff} = 39,000$~K, starlight photons have a characteristic
photon energy of $\epsilon_{\ast} \equiv h \nu_{\ast} / (m_e c^2)
\sim 7 \times 10^{-6}$, electrons and positrons with energies of
$\gamma \gtrsim 1/\epsilon_{\ast} \sim 1.5 \times 10^5$ will interact 
with these star light photons in the Klein-Nishina limit and, thus, 
very inefficiently. $\gamma$-rays produced through upscattering of
star light photons by secondary electrons and positrons in the Thomson 
regime will have energies of $E_{\rm IC} \lesssim E_{\rm IC}^{\rm max}
\equiv m_e c^2 / \epsilon_{\ast} \sim 75$~GeV. 

In addition to Compton scattering, secondary electron-positron pairs 
will suffer synchrotron losses. Magnetic fields even near the base
of microquasar jets are unlikely to exceed $B \sim 10^3$~G. Consequently,
along essentially the entire trajectory of VHE photons one may safely
assume $B \equiv 10^3 \, B_3$~G with $B_3 \lesssim 1$. Secondary electrons
and positrons traveling through such magnetic fields will produce 
synchrotron photons of characteristic energies $E_{\rm sy} \sim 1.2
\times B_3 \, \gamma_6^2$~GeV, where $\gamma_6 = \gamma / 10^6$ 
parametrizes the secondary electron's/positron's energy. Thus, even
for extreme magnetic-field values of $B = 10^3$~G, synchrotron photons
at energies $\gtrsim 100$~GeV could only be produced by secondary pairs
with $\gamma \gtrsim 10^7$, resulting from primary $\gamma$-ray photons
of $E_{\gamma} \gtrsim 10$~TeV, where only a negligible portion of the
total luminosity from LS~5039 is expected to be liberated. Consequently,
electromagnetic cascades initiated by the secondary electrons and positrons
from $\gamma\gamma$ absorption in the stellar photon field will re-emit
the absorbed VHE $\gamma$-ray photon energy essentially entirely at photon 
energies $\lesssim 100$~GeV. This conclusion is fully consistent with the 
more detailed consideration of the pair cascades resulting from $\gamma\gamma$
absorption of the VHE emission from LS~5039 by \citet{aharonian06a}.

As demonstrated in \cite{bd05}, $\gamma Z$ absorption in the stellar
wind of the companion as well as $\gamma\gamma$ absorption on star
light photons reprocessed in the stellar wind are negligible compared
to the direct $\gamma\gamma$ absorption effect for LS~5039. 

Consequently, the only relevant effect modifying the VHE $\gamma$-ray
spectrum of LS~5039 at energies above $\sim 100$~GeV between the emission
site and the observer on Earth is $\gamma\gamma$ absorption by direct 
star light photons. This effect can easily be corrected for by multiplying
the observed spectra by a factor $e^{\tau_{\gamma\gamma}(E)}$, where 
$\tau_{\gamma\gamma}(E)$ is the $\gamma\gamma$ absorption depth along 
the line of sight.

\section{\label{deabsorbed}Deabsorbed VHE $\gamma$-ray spectra}

In this section, we present a parameter study of the deabsorbed photon
spectra 
and integrated VHE $\gamma$-ray fluxes for representative values of the 
viewing angle of $i = 20^o$, $40^o$, and $60^o$ and a range of locations
of the emission site, characterized by a height $z_0$ above the compact
object in the direction perpendicular to the orbital plane. The absorption 
depth $\tau_{\gamma\gamma} (E)$ along the line of sight is evaluated as
described in detail in \cite{bd05}. The observed spectra are represented
by the best-fit functional forms quoted in the introduction. 

\begin{figure}[t]
\includegraphics[width=14cm]{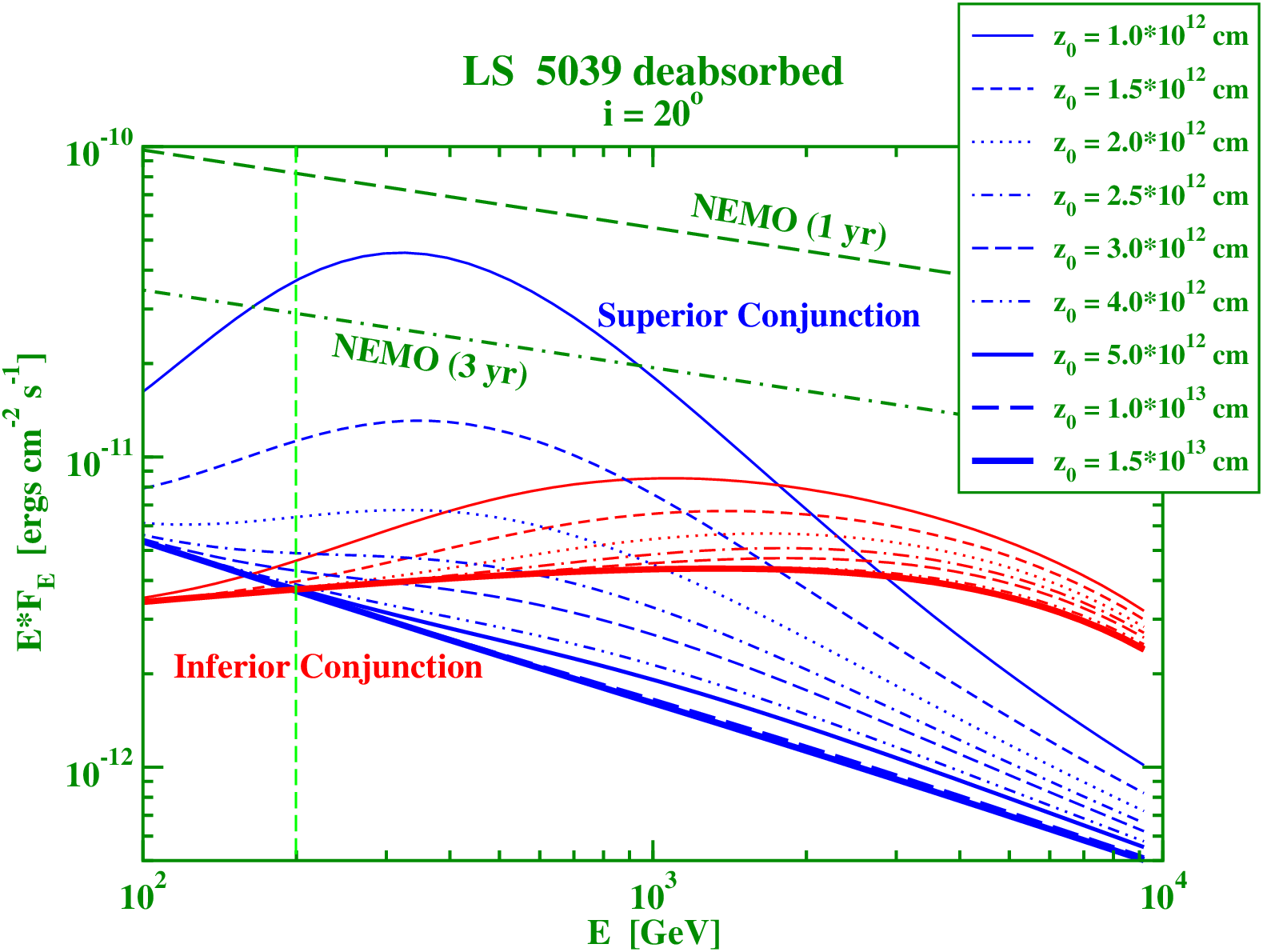}
\caption{Deabsorbed VHE $\gamma$-ray spectra of LS~5039 for $i = 20^o$ and
a range of heights of $z_0$ of the emission region above the compact object. 
The vertical dashed line indicates the energy threshold of the HESS observations.
The flat power-laws correspond to the anticipated neutrino detection sensitivities 
of the planned NEMO km$^3$ detector for a steady point source with an underlying 
power-law spectrum of neutrino number spectral index 2.5 after 1 and 3 years of
data taking, respectively \citep{distefano06}. These are relevant for comparison 
to the deabsorbed photon fluxes for a characteristic ratio of unabsorbed photons 
to neutrinos of $\sim 1$.}
\label{i20}
\end{figure}

\begin{figure}[t]
\includegraphics[width=14cm]{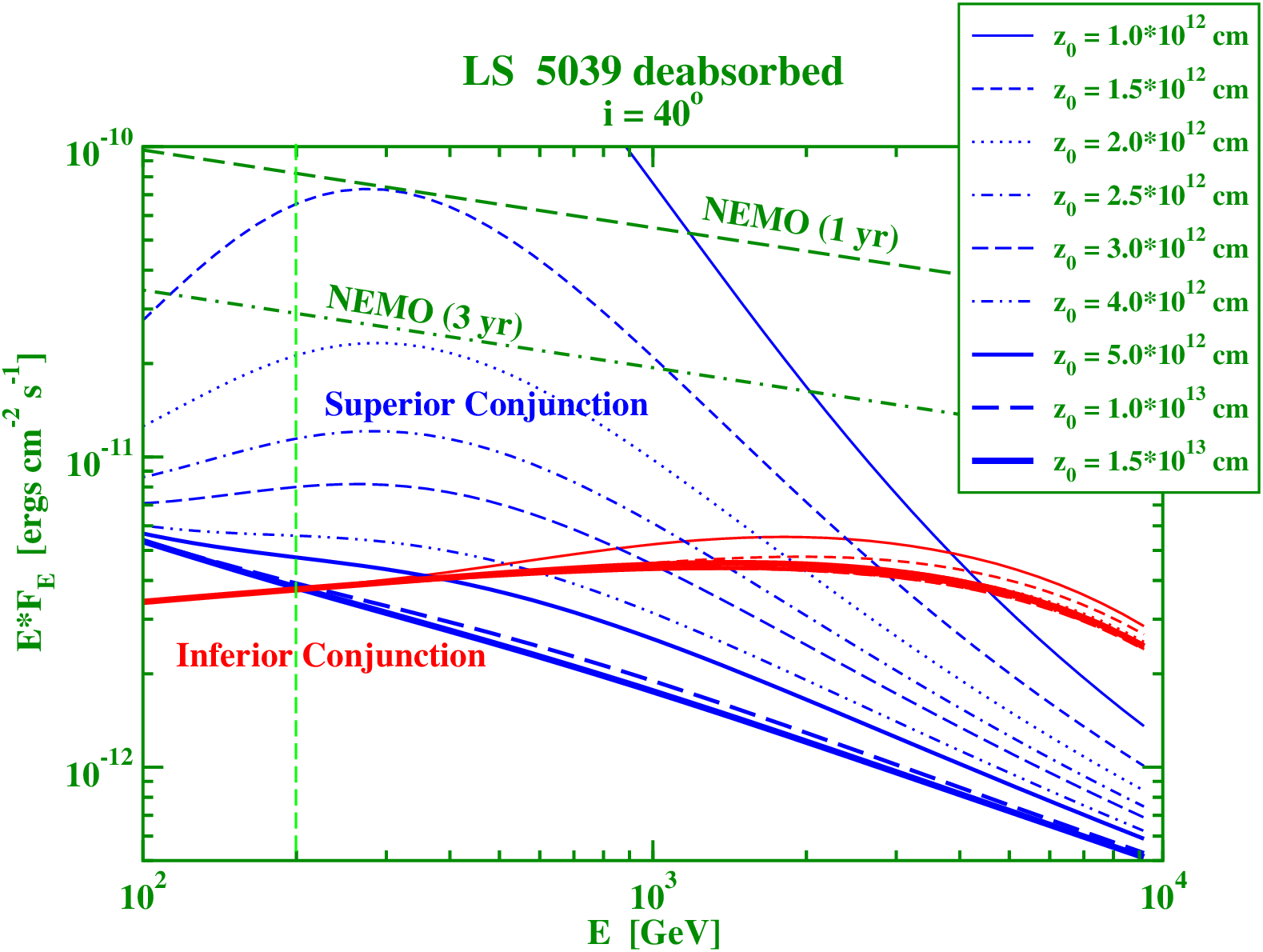}
\caption{Same as Fig. \ref{i20}, but for $i = 40^o$.}
\label{i40}
\end{figure}

\begin{figure}[t]
\includegraphics[width=14cm]{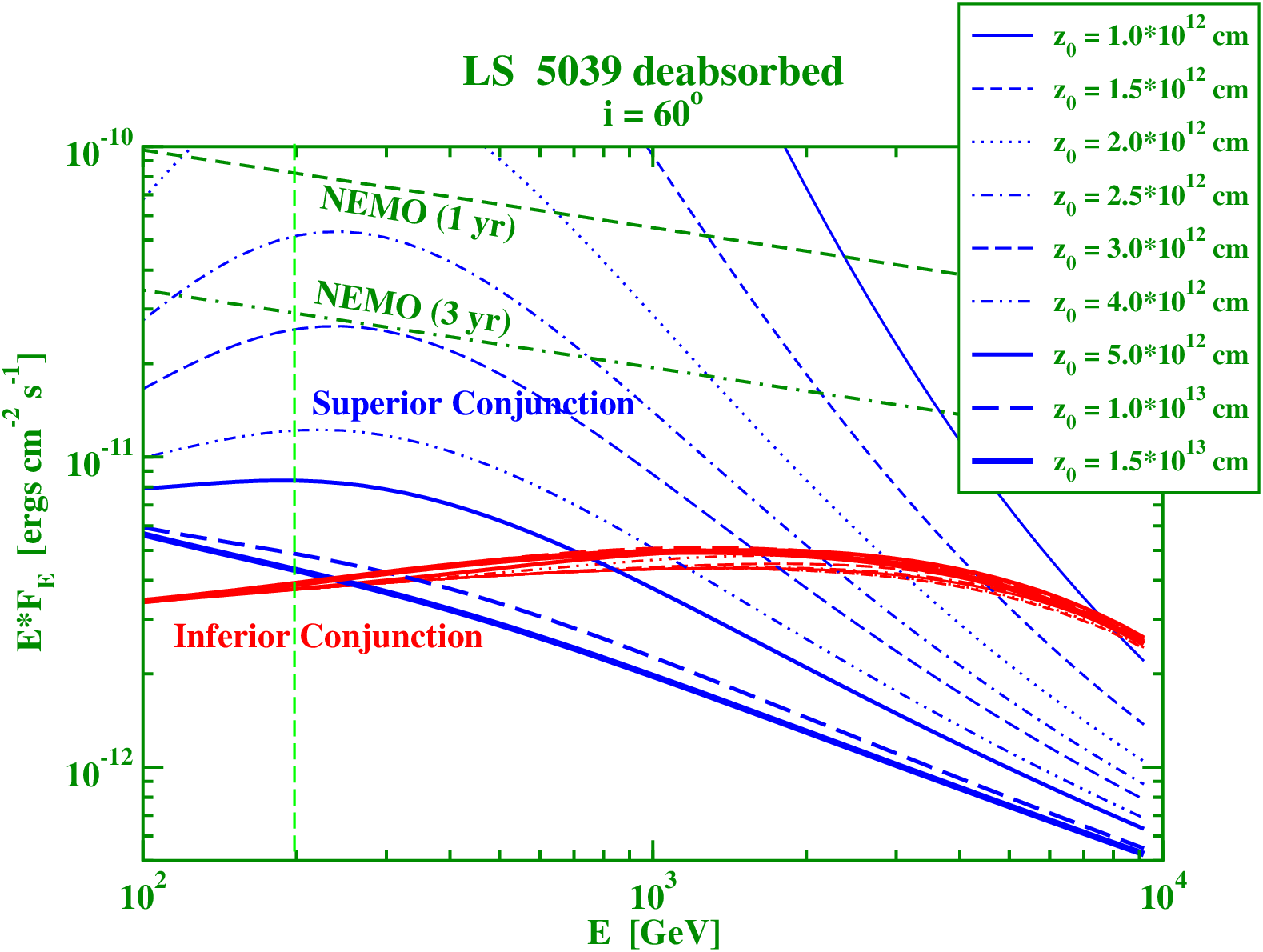}
\caption{Same as Fig. \ref{i20}, but for $i = 60^o$.}
\label{i60}
\end{figure}

Figures \ref{i20} --- \ref{i60} show the inferred intrinsic, deabsorbed
VHE $\gamma$-ray spectra from LS~5039 for a range of $10^{12} \,
{\rm cm} \le z_0 \le 1.5 \times 10^{13}$~cm. At larger distances
from the compact object, $\gamma\gamma$ absorption due to the stellar
radiation field becomes negligible for all inclination angles. 
Furthermore, models which assume a distance greatly in excess of
the characteristic orbital separation (i.e., $z_0 \gtrsim 10^{13}$~cm)
may be hard to reconcile with the periodic modulation of the emission
on the orbital time scale since the $\gamma$-ray production site
would be distributed over a large volume, with primary energy input
episodes from different orbital phases overlapping and thus smearing
out the original orbital modulation. As the
height $z_0$ approaches 
values of the order of the characteristic orbital
separation, 
$s \sim 2 \times 10^{12}$~cm, the intrinsic spectra would have
to exhibit a significant excess towards the threshold of the HESS 
observations
at $E_{\rm thr} \sim 200$~GeV in order to compensate 
for the $\gamma\gamma$
absorption trough with its extremum around 
$\sim 300$~GeV \citep{bd05}.
At even smaller distances from the 
compact object, $z_0 \ll 10^{12}$~cm, the absorption features
would again become essentially independent of $z_0$ since the overall
geometry would not change significantly anymore with a change of $z_0$.

A first, important conclusion from Figures \ref{i20} -- \ref{i60} 
is that there is no combination of $i$ and $z_0$ for which the 
de-absorbed VHE $\gamma$-ray spectra in the inferior and superior 
conjunction could be identical. Thus, the intrinsic VHE $\gamma$-ray 
spectra must be fundamentally different in the different orbital 
phases corresponding to superior conjunction (near periastron) and 
inferior conjunction (closer to apastron). 

Second, there is a large range of instances in which the deabsorbed,
differential photon fluxes at 200~GeV $\lesssim E \lesssim$ 10~TeV
during superior conjunction would have to be substantially higher than
during inferior conjunction, opposite to the observed trend. This could
have interesting consequences for strategies of searches for high-energy
neutrinos. To a first approximation, the intrinsic, unabsorbed
VHE 
$\gamma$-ray flux is roughly equal to the expected high-energy neutrino
flux \citep[see, e.g.,][]{lipari06}. Therefore, our results suggest that 
phases with lower observed VHE $\gamma$-ray fluxes from LS~5039 may actually 
coincide with phases of larger neutrino fluxes. This is confirmed by 
a plot of the integrated 0.2 -- 10~TeV photon fluxes as a function of
height of the emission region as shown in Figure \ref{flux02}. 
An anti-correlation of the VHE neutrino and $\gamma$-ray fluxes would
require emission region heights of $z_0 \lesssim 2.5 \times 10^{12}$~cm 
($i = 20^o$),
$4 \times 10^{12}$~cm ($i = 40^o$), and $7 \times 10^{12}$~cm 
($i = 60^o$),
respectively. As we will see in the next section, for 
$i = 60^o$, these configurations can be ruled out because of constraints 
on the available power from wind accretion onto the compact object. 

The possibility of a neutrino-to-photon flux ratio largely exceeding
one has also recently been discussed for the case of LSI~+61$^0$303 
by \citet{th06}. Those authors also compared the existing AMANDA upper
limits to the MAGIC VHE $\gamma$-ray fluxes of that source. They find
that a neutrino-to-photon flux ratio up to $\sim 10$ for a $\gamma\gamma$
absorption modulated photon signal would still be consistent with the
upper limits on the VHE neutrino flux from that source.

In order to assess the observational prospects of detecting VHE 
neutrinos with the new generation of km$^3$ neutrino detectors, 
such as ANTARES, NEMO, or KM3Net, one can estimate the expected 
neutrino flux assuming a characteristic ratio of produced neutrinos 
to unabsorbed VHE photons at the source of $\sim 1$ \citep{lipari06}. 
Under this assumption, the resulting neutrino fluxes should be 
comparable to the deabsorbed photon fluxes plotted in Figures 
\ref{i20} -- \ref{i60}. As a representative example of the 
sensitivity of km$^3$ water-Cherenkov neutrino detectors, those 
figures also contain the anticipated sensitivity limits of the
NEMO detector for a steady point source with a power-law of neutrino 
number spectral index $2.5$ for 1 and 3 years of data taking, respectively
\citep{distefano06}. Given the likely orbital modulation of the neutrino
flux of LS~5039, our results indicate that neutrino data accumulated over 
substantially longer than 3 years might be required in order to obtain a
firm detection of the neutrino signal from LS~5039 and its orbital modulation.

\section{\label{constraints}Constraints from Accretion Power Limits}

The inferred intrinsic (deabsorbed) VHE $\gamma$-ray fluxes calculated in 
the previous section imply apparent isotropic 0.1 -- 10~TeV luminosities 
in the range of $\sim 2 \times 10^{34}$ -- $7 \times 10^{34}$~ergs/s 
($i = 20^o$), $\sim 1.3 \times 10^{34}$ -- $7 \times 10^{35}$~ergs/s 
($i = 40^o$), and $\sim 1.2 \times 10^{34}$ -- $\gg 10^{37}$~ergs/s 
($i = 60^o$), respectively, for emission regions located at $z_0 
\ge 10^{12}$~cm (see Figures \ref{luminosities1} and \ref{luminosities2}). 
These should be confronted with constraints on the available power 
from wind accretion from the stellar companion onto the compact object.

For this purpose, we assume that for a given directed wind velocity of 
$v_{\rm wind} \approx 2.5 \times 10^8$~cm/s, all matter with $(1/2) v_{\rm wind}^2
< GM_{\rm c.o.}/R_{\rm BH}$ will be accreted onto the compact object. An 
absolute maximum on the available power is then set by $L_{\rm max} \approx 
(1/12) \, \dot M_{\rm wind} \, c^2 \, (R_{\rm BH}^2 / [4 s^2])$.  We
furthermore allow for Doppler boosting of this accretion power along the
microquasar jet with the Doppler boosting factors $D$ listed in Table 
\ref{parameter_table}. This yields an available power corresponding to an
inferred, apparent isotropic luminosity of 

\begin{equation}
L_{\rm max}^{\rm iso} \approx 2.8 \times 10^{33} \, \left( {M_{\rm c.o.} 
\over M_{\odot}} \right)^2 \, D^4 \, \left( {s \over 2 \times 10^{12} \,
{\rm cm}} \right)^{-2} \, \left( {v_{\rm wind} \over 2.5 \times 10^8
\, {\rm cm/s}} \right)^{-4} \; {\rm ergs / s}.
\label{Lmax}
\end{equation}

The resulting luminosity limits are also included in Table \ref{parameter_table}
and indicateded by the horizontal lines in Figures \ref{luminosities1} and 
\ref{luminosities2} for $\Gamma = 1.02$ and $\Gamma = 2$, respectively.
Allowed configurations
are those for which the inferred, apparent isotropic 
0.1 -- 10~TeV luminosity
is substantially below the respective luminosity 
limit. 

Additional restrictions come from the inferred luminosity of the EGRET source
3EG~J1824-1514 whose 90~\% confidence contour includes the location of LS~5039
\citep{aharonian05}. The observed $> 100$~MeV $\gamma$-ray flux from this source 
corresponds to an inferred isotropic luminosity of $L_{\rm EGRET} \sim 7 \times 
10^{34}$~ergs~s$^{-1}$. This level is also indicated in Figures \ref{luminosities1} 
and \ref{luminosities2}. However, the association of the EGRET source with LS~5039 
is uncertain since the 90~\% confidence contour also contains the pulsar
PSR~B1822-14 as a possible counterpart. Thus, even if the available power 
according to eq. \ref{Lmax} is lower than $L_{\rm EGRET}$, this would not
provide a significant problem since the EGRET flux could be provided by the 
nearby pulsar. However, the EGRET flux does provide an upper limit on the 
$\sim 30$~MeV -- 30~GeV flux from LS~5039. As shown in \citet{aharonian06a},
the bulk of the VHE $\gamma$-ray flux that is absorbed by $\gamma\gamma$
absorption on companion star photons and initiates pair cascades, will be
re-deposited in $\gamma$-rays in the EGRET energy range. Thus, any scenario
that requires an unabsorbed source luminosity in excess of the EGRET luminosity
would be very problematic. The resulting limits inferred from both eq. \ref{Lmax}
and the EGRET flux on the height $z_0$ are summarized in Table \ref{z_limits}.

The figure indicates that for $i = 20^o$, virtually all configurations are 
allowed, even if the emission is essentially
unboosted ($\Gamma = 1.02$)
and originates very close to the compact object. 

For $i = 40^o$ at the time of superior conjunction, one can set a limit 
of $z_0 \gtrsim 1.7 \times 10^{12}$~cm. For this viewing angle at inferior 
conjunction, the limit implied for $\Gamma = 1.02$ is always below the 
inferred value, indicating that this configuration is ruled out. This 
implies that, if $i \sim 40^o$, the emission at the time of inferior 
conjunction must be substantially Doppler enhanced.

The situation is even more extreme for $i = 60^o$. Under this viewing angle,
only the $\Gamma = 1.02$ scenario leads to an allowed model and requires $z_0
\gtrsim 5 \times 10^{12}$~cm. Since there is no allowed scenario to produce 
the observed flux at inferior conjunction, one can
conclude that a large 
inclination angle of $i \sim 60^o$ may be ruled out.

Note, however, that there are alternative models for the very-high-energy 
emission from X-ray binaries in which the source of power for the high-energy 
emission lies in the rotational energy of the compact object \citep[in that 
case, most plausibly a neutron star, see, e.g.][]{chernyakova06}.
The constraints discussed above will obviously not apply to such models. 
Instead, in the case of a rotation-powered pulsar wind/jet, one can estimate 
the total available power as

\begin{equation}
L_{\rm rot} \sim {8 \over 5} \pi^2 \, M \, R^2 \, {\dot P \over P^3} \sim
4.4 \cdot 10^{37} \, {\dot P_{-15} \over P_{-2}^3} \; {\rm ergs \; s}^{-1},
\label{L_rotation}
\end{equation}
where a $1.4 \, M_{\odot}$ neutron star with $R = 10$~km is assumed and the
spin-down rate and spin period are parametrized as $\dot P = 10^{-15} \, 
\dot P_{-15}$ and $P = 10 \, P_{-2}$~ms, respectively. Consequently, the
power requirements of even the extreme scenarios for $i = 60^o$, as discussed 
above, could, in principle, be met by a rotation-powered pulsar. However, 
such a luminosity could only be sustained over a spin-down time of

\begin{equation}
t_{\rm sd} \sim 3 \times 10^5 \, {P_{-2} \over \dot P_{-15}} \; {\rm years}
\label{t_spindown}
\end{equation}
and should therefore be rare.

\section{\label{summary}Summary and Discussion}

\begin{figure}[t]
\includegraphics[width=14cm]{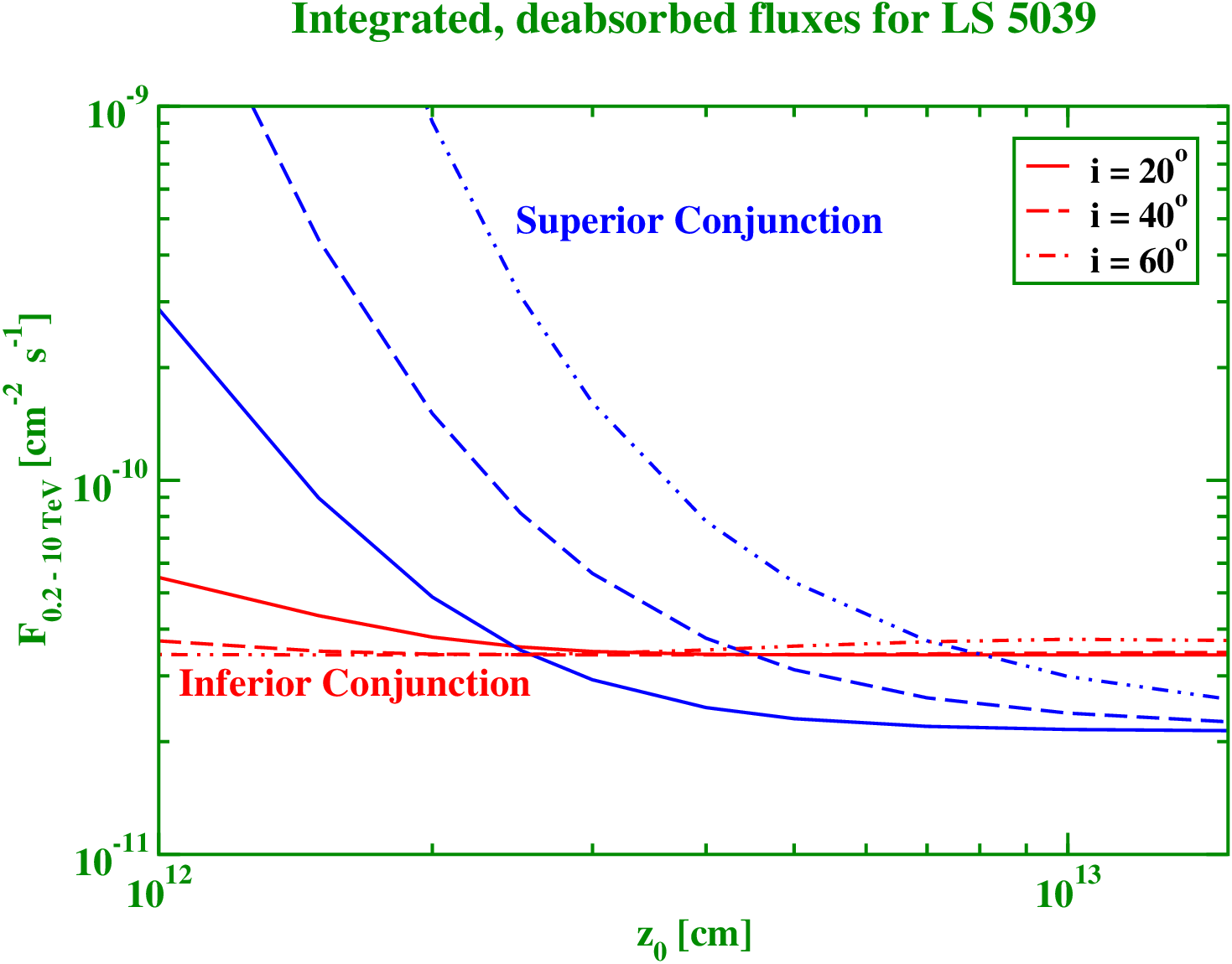}
\caption{Deabsorbed, integrated 0.2 -- 10~TeV $\gamma$-ray fluxes of LS~5039 
as a function of height $z_0$ of the emission region above the compact object. 
These numbers would also approximately equal to the expected neutrino fluxes
in the same energy range.}
\label{flux02}
\end{figure}

\begin{figure}[t]
\includegraphics[width=14cm]{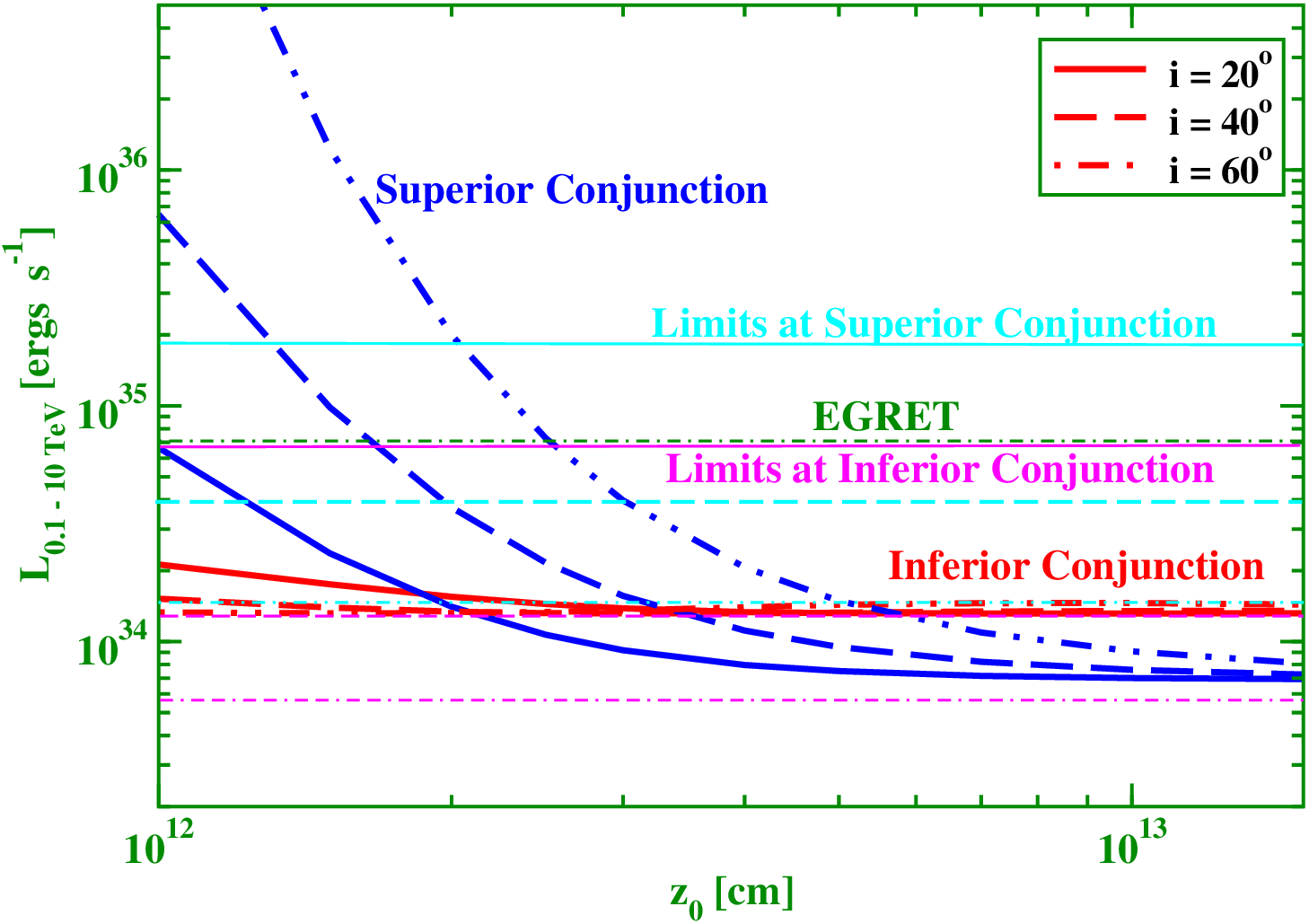}
\caption{Inferred apparent isotropic luminosities for LS~5039 (thick
curves), compared to the luminosity limits from Bondi-Hoyle limited
wind accretion (horizontal lines), assuming Doppler boosting of the
emission corresponding to $\Gamma = 1.02$. Allowed configurations 
are those for which the inferred
luminosities are substantially below 
the respective luminosity limits.}
\label{luminosities1}
\end{figure}

\begin{figure}[t]
\includegraphics[width=14cm]{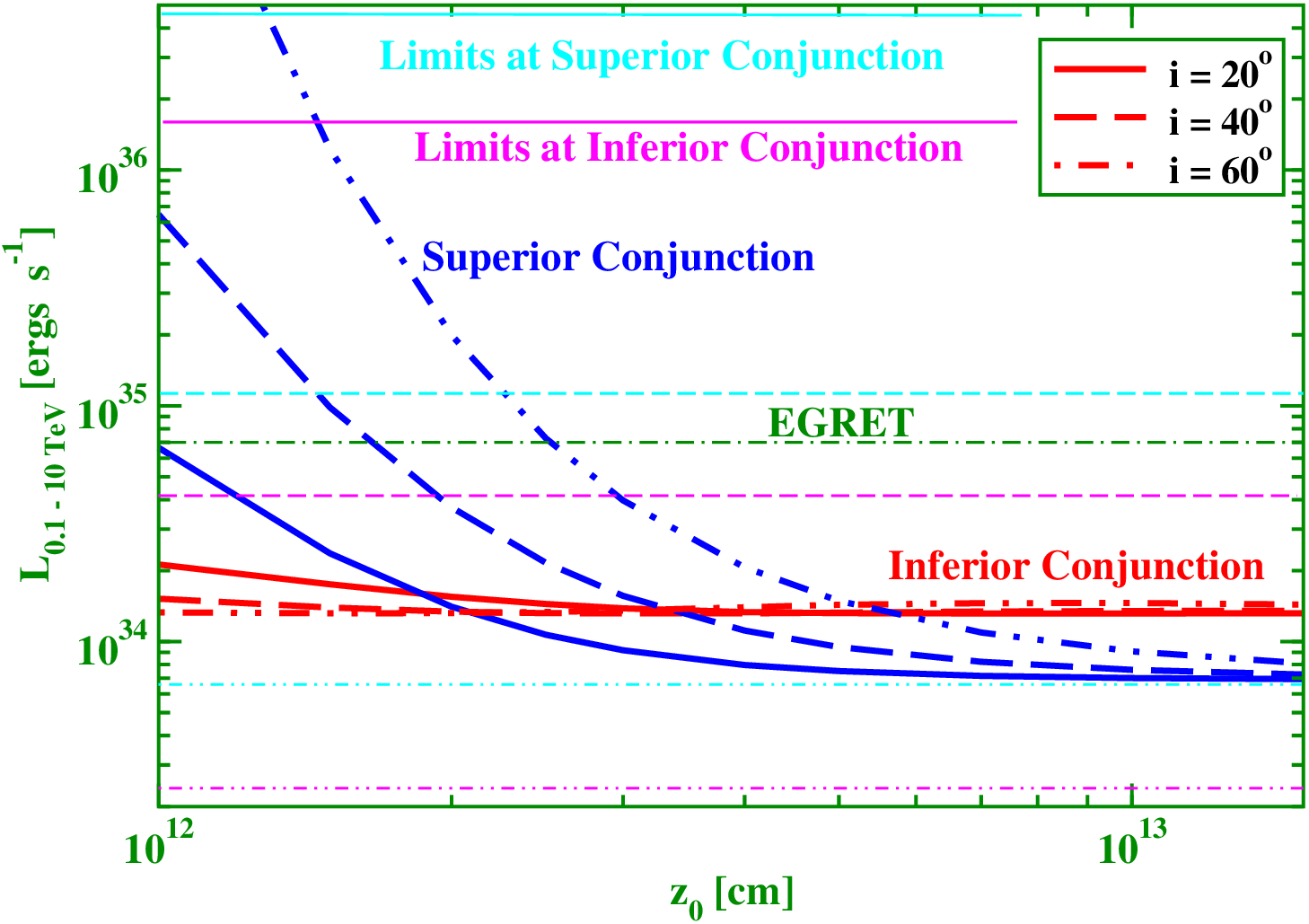}
\caption{Same as Fig. 6, but for $\Gamma = 2$.}
\label{luminosities2}
\end{figure}

In this paper, a detailed study of the intrinsic VHE $\gamma$-ray emission from 
the Galactic microquasar LS~5039 after correction for $\gamma\gamma$ absorption
by star light photons from the massive companion star is presented. This
system had shown evidence for an orbital modulation of the VHE $\gamma$-ray
flux and spectral shape. A range of observationally allowed inclination angles,
$13^o \le i \le 64^o$ (specifically, $i = 20^o$, $i = 40^o$, and $i = 60^o$)
as well as plausible distances $z_0$ of the VHE $\gamma$-ray emission
region above the compact object ($10^{12} \, {\rm cm} \le z_0 \le 1.5 \times
10^{13}$~cm) was explored. Deabsorbed, intrinsic VHE $\gamma$-ray spectra as 
well as integrated fluxes and inferred, apparent isotropic luminosities were 
calculated and contrasted with constraints from the available power from 
wind accretion from the massive companion onto the compact object. The main 
results are:

\begin{itemize}

\item{It is impossible to choose the viewing angle and location of the VHE 
emission region in a way that the intrinsic (deabsorbed) fluxes and spectra 
in superior and inferior conjunction are identical within the range of values
of $i$ and $z_0$ considered here. For values of $z_0$ much smaller and much
larger than the characteristic orbital separation ($s \sim 2 \times 10^{12}$~cm), 
the absorption features would become virtually independent of $z_0$. Furthermore,
models assuming a VHE $\gamma$-ray emission site at a distance greatly in excess
of the orbital separation might be difficult to reconcile with the orbital 
modulation of the VHE emission. Consequently, the intrinsic VHE luminosities 
and spectral shapes must be fundamentally different in different 
orbital phases.}

\item{It was found that the luminosity constraints for an inclination angle
of $i = 60^o$ at inferior conjunction could not be satisfied at all, and
for $i = 40^o$, there is no allowed configuration in agreement with the
luminosity constraint for $\Gamma = 1.02$. From this, it may be concluded
that, if the VHE luminosity is limited by wind accretion from the companion
star and the system is viewed at an inclination angle of $i \gtrsim 40^o$, 
the emission is most
likely beamed by a larger Doppler factor than 
inferred from the dynamics of the large-scale radio outflows on scales
of several hundred AU.}

\item{Since it was found to be impossible to satisfy the luminosity constraint
for inferior conjunction at a viewing angle of $i = 60^o$, one can constrain
the viewing angle to values substantially smaller than the maximum of $\sim 64^o$ 
inferred from the lack of eclipses.}

\item{The previous two points as well as the fact that the luminosity limits
can easily be satisfied for $i = 20^o$, indicate that a rather small inclination 
angle $i \sim 20^o$ may be preferred. Thus, our results confirm the conjecture 
of \cite{casares05} that the compact
object might be a black hole rather than 
a neutron star.}

\item{Under the assumption of a photon-to-neutrino ratio of $\sim 1$ 
(before $\gamma\gamma$ absorption), the detection of the neutrino flux 
from LS~5039 and its orbital modulation might require the accumulation
of data over more than 3 years with km$^3$ scale neutrino detectors like
ANTARES, NEMO, or KM3Net.}

\item{Comparing the ranges of allowed configurations from Table \ref{z_limits}
to the plot of intrinsic VHE fluxes in Figure \ref{flux02}, one can 
see that there is a limited range of allowed configurations for which the 
expected VHE neutrino flux would actually anti-correlate with the observed 
VHE $\gamma$-ray emission. Specifically, for a preferred viewing angle of 
$i \sim 20^o$, models in which the emission originates within $z_0 \lesssim 
2.5 \times 10^{12}$~cm would predict that the VHE neutrino flux at superior 
conjunction is larger than at inferior conjunction, opposite to the orbital 
modulation trend seen in VHE $\gamma$-ray photons. Thus, strategies for the 
identification of high-energy neutrinos from microquasars based on a positive 
correlation with observed VHE fluxes may fail if models with $z_0 \lesssim 
2.5 \times 10^{12}$~cm apply.}

\item{The luminosity limitations discussed above could, in principle, be 
overcome by a rotation-powered pulsar. However, this would require a rather 
short spin-down time scale of $t_{\rm sd} \lesssim 10^6$~yr. Consequently, such 
objects should be rare.}

\end{itemize}

\acknowledgments
The author thanks F. Aharonian and V. Bosch-Ramon for stimulating
discussions and hospitality during a visit at the Max-Planck-Institute
for Nuclear Physics, and M. De Naurois for providing the HESS data 
points. I also thank the referee for very insightful and constructive
comments, which have greatly helped to improve this manuscript. 
This work was partially supported by NASA INGEGRAL Theory 
grant no. NNG~05GP69G and a scholarship at the Max-Planck Institute 
for Nuclear Physics in Heidelberg.

\newpage

\newpage

\newpage

\newpage

\newpage

\newpage

\newpage

\newpage

\begin{deluxetable}{ccccccc}
\tabletypesize{\scriptsize}
\tablecaption{Inferred parameters of the LS~5039 system for three
representative values of the inclination angle.}
\tablewidth{0pt}
\tablehead{
\colhead{$i$ [$^o$]} & 
\colhead{position} & 
\colhead{$M_{\rm c.o.}$ [$M_{\odot}$]} & 
\colhead{$D (\Gamma = 1.02)$} & 
\colhead{$D (\Gamma = 2)$} &
\colhead{$L_{\rm max}^{\rm iso} (\Gamma = 1.02)$ [erg/s]} &
\colhead{$L_{\rm max}^{\rm iso} (\Gamma = 2)$ [erg/s]}
}
\startdata
20 & i.c. & 4.5 & 1.21 & 2.68 & $6.7 \times 10^{34}$ & $1.6 \times 10^{36}$ \\
20 & s.c. & 4.5 & 1.21 & 2.68 & $1.9 \times 10^{35}$ & $4.6 \times 10^{36}$ \\
40 & i.c. & 2.3 & 1.13 & 1.49 & $1.3 \times 10^{34}$ & $4.0 \times 10^{34}$ \\
40 & s.c. & 2.3 & 1.13 & 1.49 & $3.8 \times 10^{34}$ & $1.1 \times 10^{35}$ \\
60 & i.c. & 1.6 & 1.09 & 0.88 & $5.6 \times 10^{33}$ & $2.4 \times 10^{33}$ \\
60 & s.c. & 1.6 & 1.09 & 0.88 & $1.6 \times 10^{34}$ & $6.7 \times 10^{33}$ \\
\enddata
\label{parameter_table}
\end{deluxetable}

\newpage

\begin{deluxetable}{cccc}
\tabletypesize{\scriptsize}
\tablecaption{Lower limits on the height of the VHE $\gamma$-ray emission site
above the compact object inferred from Figures \ref{luminosities1} and 
\ref{luminosities2}. Configurations for which the intrinsic apparent isotropic 
luminosity would always be larger than the available accretion power can not 
be realized and are marked as ``forbidden''.}
\tablewidth{0pt}
\tablehead{
\colhead{$i$ [$^o$]} & 
\colhead{position} & 
\colhead{$z_0^{\rm min} (\Gamma = 1.02)$ [$10^{12}$ cm]} & 
\colhead{$z_0^{\rm min} (\Gamma = 2)$ [$10^{12}$ cm]}
}
\startdata
20 & i.c. & $\ll 1$ & $\ll 1$ \\
20 & s.c. & $\ll 1$ & $\ll 1$ \\
40 & i.c. & forbidden & $\ll 1$ \\
40 & s.c. & $2.0$ & $1.7$ \\
60 & i.c. & forbidden & forbidden \\
60 & s.c. & 5 & forbidden \\
\enddata
\label{z_limits}
\end{deluxetable}


\begin{thebibliography}

\bibitem[Aharonian et al.(2005)]{aharonian05}Aharonian, F. A., et al.,
2005, Science, 309, 746

\bibitem[Aharonian et al.(2006a)]{aharonian06a}Aharonian, F. A., Anchordoqui, L. A.,
Khangulyan, D., \& Montaruli, T., 2006a, in proc. of 9th International Conference
on Topics in Astrophysics and Underground Physics, "TAUP 2005", Journal of Physics,
Conf. Series, in press (astro-ph/0508658)

\bibitem[Aharonian et al.(2006b)]{aharonian06b}Aharonian, F. A., et al., 
2006b, A\&A, submitted (astro-ph/0607192)

\bibitem[Albert et al.(2006)]{albert06}Albert, J., et al., 2006,
Science, vol. 312, issue 5781, p. 1771

\bibitem[Bednarek(1997)]{bednarek97}Bednarek, W., 1997, A\&A, 322, 523

\bibitem[Bednarek(2000)]{bednarek00}Bednarek, W., 2000, A\&A, 362, 646

\bibitem[B\"ottcher \& Dermer(2005)]{bd05} B\"ottcher, M., \& Dermer,
C. D., 2005, ApJ, 634, L81

\bibitem[Bosch-Ramon \& Paredes(2004)]{bp04}Bosch-Ramon, V., \& Paredes,
J. M., 2004, A\&A, 417, 1075

\bibitem[Bosch-Ramon et al.(2005a)]{bosch05a}Bosch-Ramon, V., Romero, G. E., 
\& Paredes, J. M., 2005a, A\&A, 429, 267

\bibitem[Bosch-Ramon et al.(2005b)]{bosch05b}Bosch-Ramon, V., Aharonian, F. A.,
\& Paredes, J. M., 2005b, A\&A, 432, 609

\bibitem[Casares et al.(2005)]{casares05}Casares, J., Rib\'o, M.,
Ribas, I., Paredes, J. M., Mart\'i, J., \& Herrero, A., 2005, MNRAS,
364, 899

\bibitem[Chernyakova, Neronov, \& Walter(2006)]{chernyakova06}
Chernyakova, M., Neronov, A., \& Walter, R., 2006, MNRAS, in press
(astro-ph/0606070; early release under 2006MNRAS.tmp.1051C)

\bibitem[Dermer \& B\"ottcher(2006)]{db06}Dermer, C. D., \& B\"ottcher, M.,
2006, ApJ, 643, 1081

\bibitem[Distefano(2006)]{distefano06}Distefano, C., 2006, Astroph. \& Space Science,
submitted (astro-ph/0608514)

\bibitem[Gregory(2002)]{gregory02}Gregory, P. C., 2002, ApJ, 575, 427

\bibitem[Dubus(2006a)]{dubus06a}Dubus, G., 2006a, A\&A, 451, 9

\bibitem[Dubus(2006b)]{dubus06b}Dubus, G., 2006b, A\&A, 456, 801

\bibitem[Gregory \& Taylor(1978)]{gregory78}Gregory, P. C., \& Taylor, A. R.,
1978, Nature, 272, 704

\bibitem[Gupta, B\"ottcher, \& Dermer(2006)]{gupta06}Gupta, S., B\"ottcher, M.,
\& Dermer, C. D., 2006, ApJ, 644, 409

\bibitem[Gupta \& B\"ottcher(2006)]{gb06}Gupta, S., \& B\"ottcher, M., 2006,
ApJ, ApJ, 650, L123

\bibitem[Kniffen et al.(1997)]{kniffen97} Kniffen, D.~A., et al.\ 1997, ApJ, 
486, 126 

\bibitem[Lipari(2006)]{lipari06}Lipari, P., in proc. of ``Very Large Volume
Neutrino Telescopes'', Catania, Italy, 2005, in press (astro-ph/0605535)

\bibitem[McSwain \& Gies(2002)]{mg02}McSwain, M. V., \& Gies, D. R., 2002, 
ApJ, 568, L27

\bibitem[Moskalenko et al.(1993)]{moskalenko93}Moskalenko, I. V., Karakula, S.,
\& Tkaczyk, W., 1993, MNRAS, 260, 681

\bibitem[Moskalenko \& Karakula(1994)]{moskalenko94}Moskalenko, I. V., \&
Karakula, S., 1994, ApJS, 92, 567

\bibitem[Paredes et al.(2000)]{paredes00}Paredes, J. M., Mart\'\i, J., Rib\'o, M.,
\& Massi, M., 2000, Science, 288, 2340

\bibitem[Paredes et al.(2002)]{paredes02}Paredes, J. M., Rib\'o, M., Ros, E.,
Mart\'\i, J., \& Massi, M., 2002, A\&A, 393, L99

\bibitem[Reimer et al.(2006)]{reimer06}Reimer, A., Pohl, M., \& Reimer, O.,
2006, ApJ, 644, 1118

\bibitem[Romero et al.(2003)]{romero03}Romero, G. E., Torres, D. F., Kaufman Bernad\'o,
M. M., \& Mirabel, I. F., 2003, A\&A, 410, L1

\bibitem[Taylor et al.(1992)]{taylor92}Taylor, A. R., Kenny, H. T. Spencer, R. E.,
\& Tzioumis, A., 1992, ApJ, 395, 268

\bibitem[Torres \& Halzen(2006)]{th06}Torres, D., \& Halzen, F., 2006, A\&A, submitted
(astro-ph/0607368)

\end{thebibliography}
\end{document}